\documentclass[onecolumn,aps]{revtex4-2}
\usepackage{orcidlink}
\usepackage{mathrsfs}
\usepackage{physics}
\usepackage{amsfonts}
\usepackage{dcolumn}
\usepackage{bm}
\usepackage{amssymb}
\usepackage{pgf}
\usepackage{fancyhdr}
\usepackage{amsmath}
\usepackage{braket}
\usepackage{amsthm}
\usepackage{epsfig}
\usepackage{epstopdf}
\usepackage{graphicx}
\usepackage{hyperref}
\usepackage{booktabs}
\usepackage{xcolor}
\usepackage[title]{appendix}
\usepackage{cleveref}
\hypersetup{
colorlinks=true,
linkcolor=blue,
urlcolor=blue,
citecolor=blue
}

\newtheorem{theorem}{Theorem}

\newtheorem{corollary}{Corollary}
\newtheorem{example}{Example}

\begin{document}
\title{Separability and lower bounds of concurrence via symmetric measurement correlations}
\author{Yuhang Xie}
\affiliation{School of Mathematics and Statistics, Henan University, Kaifeng 475004, China}
\author{Yanjun Chu}
\email{chuyj@henu.edu.cn}
\affiliation{School of Mathematics and Statistics, Henan University, Kaifeng 475004, China}
\author{Shao-Ming Fei}
\affiliation{School of Mathematics Sciences, Capital Normal University, Beijing 100048, China}
\email{feishm@cnu.edu.cn}
\begin{abstract}
The detection and quantification of quantum entanglement is a central issue in quantum information science, particularly for unknown quantum states. Based on symmetric measurement correlation matrices, we present a family of separability criteria that can be experimentally applied without state tomography, and recover some related existing criteria as special cases. Using the constructed correlation matrices, we further derive a class of lower bounds of concurrence. Through illustrative examples, we demonstrate that our criteria outperform some existing separability criteria in entanglement detection, and that the obtained lower bounds on concurrence are tighter than known counterparts.
\end{abstract}

\maketitle

\section{Introduction}\label{sec1}
Quantum entanglement constitutes a fundamental resource for quantum information processing. Nevertheless, the detection and quantification of entanglement remain a core challenge. A bipartite state \(\rho_{AB}\) is separable if it can be expressed as a convex combination of product states,
\begin{equation*}
\rho_{AB} = \sum_k p_k \rho^A_k \otimes \rho^B_k, \quad p_k \geq 0 ,~~ \sum_k p_k=1,
\end{equation*}
where $ \rho^A_k$ ($\rho^B_k$) are density matrices of subsystems $A$ ($B$). In general, there is no operational criterion capable of detecting all entangled states. Commonly adopted necessary conditions for separability encompass the positive partial transpose (PPT) criterion \cite{PeresPRL1413,HorodeckiPLA1996,HorodeckiPLA333}, the realignment criterion \cite{RudolphPRA032312,ChenWuQIC193,HorodeckiOSID103,
AlbevrioChenFeiPRA062313}, the covariance matrix criterion \cite{GuhnePRL130504}, the correlation matrix criterion \cite{VicenteQIC624,VicenteJPA065309,LiMingPRA022325}, and the moment-based criterion \cite{ElbenPRL200501,YuPRL060504,ZhangQIP276,WangEPJP1378,AggarwalPRA012404,ZhangQIP238,ZhaoQIP90,MukherjeePRA062428}. 

Measurement-based criteria are particularly appealing, as they rely exclusively on local measurement statistics and obviate the need for full state tomography. Accordingly, measurement-based separability criteria furnish an operational framework for entanglement detection, drawing on outcome statistics obtained from feasible local positive operator-valued measures (POVMs). Spengler \textit{et al}. \cite{Spengler2012} employed mutually unbiased bases (MUBs) \cite{DurtIJQI535} to detect quantum entanglement across various quantum systems. Nevertheless, it remains an open question whether a complete set of MUBs exists in dimensions that are not prime powers. To circumvent this limitation, Kalev and Gour proposed mutually unbiased measurements (MUMs) which can be constructed in arbitrary finite dimensions~\cite{KalevGour2014}. In Ref.~\cite{ChenMaFei2014}, the authors developed a MUM-based separability criterion applicable to arbitrary bipartite systems. In parallel, symmetric informationally complete POVMs (SIC-POVMs) offer an alternative path toward measurement-based separability criteria. Renes \textit{et al}. laid the foundational framework for SIC-POVMs in Ref.~\cite{Renes2004JMP}, while Shang \textit{et al}. constructed an improved entanglement criterion using SIC-POVMs \cite{Shang2018}. General SIC-POVMs (GSIC-POVMs) were introduced in Ref.~\cite{Appleby2007} to generalize the SIC-POVM constructions beyond rank-one measurement operators. Later, Gour and Kalev \cite{GourKalev2014} presented a construction scheme for the corresponding measurement operators using Gell-Mann matrices. Following these advances, numerous separability criteria based on GSIC-POVMs have been proposed \cite{ChenLiFei2015,Lai2018,XiZhengZhuQIP2016}. Recently, the framework of $(N,M)$-POVMs was introduced by Siudzi\'nska in Ref.~\cite{SiudzinskaPRA2022042209}, which includes MUMs and GSIC-POVMs as special cases. Using joint outcome probabilities from local \((N,M)\)-POVMs, Lu \textit{et al.} proposed a bipartite separability criterion and outlined its generalization to multipartite scenarios~\cite{LuPRA062424}. Later, an enhanced version of this criterion was derived via a more flexible parameterized correlation matrix~\cite{LuAQTe70344}. These works establish the usefulness of the symmetric-measurement framework. However, the auxiliary upper-left blocks of their constructed matrix used for entanglement detection are outer product and hence have rank at most one. The resulting rank-one structure limits how much matrix-valued probability information this block can encode. In this work, we generalize this construction by allowing the upper-left block to have arbitrary rank. This extension unifies and strengthens the aforementioned criteria.

Besides entanglement detection, the quantification of entanglement represents another core topic in quantum information theory. Among diverse entanglement measures, the concurrence has attracted extensive attention \cite{RungtaPRA042315,ChenAlbeverioFei}. Nevertheless, for an arbitrary mixed bipartite state \(\rho\), the concurrence \(C(\rho)\) is defined via the convex-roof extension, which renders its exact evaluation intractable except for special quantum states. The framework of symmetric measurements has also been exploited for quantitative estimation of concurrence. In Ref.~\cite{WangFeiConcurrence}, a symmetric-measurement-induced lower bound of concurrence was derived without requiring full state tomography. Follow-up work refined this bound by introducing a correlation matrix construction~\cite{WangEtAlEstimatingConcurrence}. In this work, we derive a new lower bound of concurrence by using the generalized correlation matrix developed in the present paper, which covers the related lower bounds as special cases. This result extends the symmetric-measurement construction strategy to the adjustable-rank family considered herein. We benchmark the proposed bound against existing ones through several illustrative examples.

The remainder of this paper is organized as follows. Section~\ref{sec2} reviews symmetric $(N,M)$-POVMs and the associated probability matrices. Section~\ref{sec3} presents the generalized bipartite entanglement criterion, together with its special cases and illustrative examples. Section~\ref{sec:concurrence} derives the concurrence lower bound and discusses examples. Finally, we conclude in Sec.~\ref{sec5}.

\section{Preliminaries}\label{sec2}

The symmetric \((N,M)\)-POVMs proposed in Ref.~\cite{SiudzinskaPRA2022042209} consist of $N$ $d$-dimensional POVMs given by measurement operators \(\{E_{\alpha,k}\mid k=1,2,\dots,M\}\) with \(\alpha=1,2,\dots,N\). Each POVM contains $M$ measurement operators satisfying
\begin{equation}\label{NM-POVMs-Definition}
\begin{aligned}
\mathrm{Tr}(E_{\alpha,k})&=\frac{d}{M},\\[2pt]
\mathrm{Tr}\left(E_{\alpha,k}^{2}\right)&=x,\\[2pt]
\mathrm{Tr}\left(E_{\alpha,i}E_{\alpha,j}\right)&=\frac{d-Mx}{M(M-1)}\quad(i \neq j),\\[2pt]
\mathrm{Tr}\left(E_{\alpha,k}E_{\beta,l}\right)&=\frac{d}{M^{2}}\quad(\alpha\neq\beta),
\end{aligned}
\end{equation}
where the parameter $x$ obeys the constraint $\frac{d}{M^{2}}<x\leq\min\left\{\frac{d^{2}}{M^{2}},\frac{d}{M}\right\}$. In particular, \((N,M)\)-POVMs reduce to projective measurements for \(x=\frac{d^{2}}{M^{2}}\), which implies that projective \((N,M)\)-POVMs exist only when \(M \geq d\). An \((N,M)\)-POVM is termed informationally complete if \(N(M-1)=d^{2}-1\).

Let $\{G_{0}=I_{d}/\sqrt{d},\,G_{\alpha,k}\mid \alpha=1,\dots,N;\,k=1,\dots,M-1\}$ be an orthonormal basis for the space of $d\times d$ Hermitian operators, satisfying $\mathrm{Tr}(G_{\alpha,k})=0$, where $I_d$ denotes the $d\times d$ identity operator. The measurement operators of an informationally complete $(N,M)$-POVM have the following explicit expressions,
\begin{equation}\label{NM-POVMs-Construction}
E_{\alpha,k}=\frac{1}{M}I_{d}+tH_{\alpha,k},
\end{equation}
where
\begin{equation*}
H_{\alpha,k}=
\begin{cases}
G_{\alpha}-\sqrt{M}\left(\sqrt{M}+1\right)G_{\alpha,k}, & k=1,\cdots,M-1,\\[2pt]
\left(\sqrt{M}+1\right)G_{\alpha}, & k=M,
\end{cases}
\end{equation*}
with $G_{\alpha}=\sum\limits_{k=1}^{M-1}G_{\alpha,k}$. The positivity of $E_{\alpha,k}$ imposes the condition
\begin{equation}\label{range-of-t}
-\frac{1}{M}\frac{1}{\lambda_{\max}}\leq t \leq\frac{1}{M}\frac{1}{|\lambda_{\min}|},
\end{equation}
where $\lambda_{\max}$ and $\lambda_{\min}$ denote, respectively, the maximum and minimum eigenvalues of all $H_{\alpha,k}$. The parameters $t$ and $x$ satisfy the relation
\begin{equation}\label{t-and-x-Relation}
x=\frac{d}{M^{2}}+t^{2}(M-1)\left(\sqrt{M}+1\right)^{2}.
\end{equation}

For an \((N,M)\)-POVMs \(\{\{E_{\alpha,k}\}_{k=1}^M\}_{\alpha=1}^N\), it has been shown in Ref.~\cite{SiudzinskaPRA2022042209} that the following identity holds for any state $\rho$,
\begin{equation}\label{sum-of-the-squares-of-probabilities}
C(x,\rho)=\sum_{\alpha=1}^{N}\sum_{k=1}^{M}\langle E_{\alpha,k}\rangle_{\rho}^{2}
=\frac{d(M^{2}x-d)\mathrm{Tr}(\rho^{2})+d^{3}-M^{2}x}{dM(M-1)},
\end{equation}
where \(\langle E_{\alpha,k}\rangle_{\rho}=\mathrm{Tr}(E_{\alpha,k}\rho)\). Therefore, the condition $\mathrm{Tr}(\rho^{2})\leq1$ implies that $C(x,\rho)$ has the following upper bound,
\begin{equation}\label{upper-bound-of-the-sum-of-the-squares-of-probabilities}
C(x,\rho)\leq\frac{d-1}{d}\frac{M^{2}x+d^{2}}{M(M-1)}.
\end{equation}

Two special families of \((N,M)\)-POVMs are of particular interest.

1) {\bf GSIC-POVMs.} If \(N=1\) and \(M=d^{2}\), the \((N,M)\)-POVMs reduce to GSIC-POVMs \cite{Appleby2007,GourKalev2014}, consisting of \(d^{2}\) positive semi-definite operators \(\{M_{k}\}_{k=1}^{d^{2}}\) that satisfy
\begin{equation}\label{GSICdef}
\begin{split}
\mathrm{Tr}(M_{k})&=\frac{1}{d},\\[2pt]
\mathrm{Tr}\left(M_{k}^{2}\right)&=a,\\[2pt]
\mathrm{Tr}(M_{j}M_{k})&=\frac{1-da}{d(d^{2}-1)}\quad(j \neq k),
\end{split}
\end{equation}
where the parameter $a$ satisfies $\frac{1}{d^3}<a\leq\frac{1}{d^{2}}$. When $a=\frac{1}{d^{2}}$, the operators $M_{k}$ are rank-one projectors and the GSIC-POVMs become SIC POVMs. Let $\{F_{k}\}_{k=1}^{d^{2}-1}$ be a set of $d^{2}-1$ Hermitian and traceless operators on $\mathbb{C}^{d}$ such that $\mathrm{Tr}(F_{k}F_{l})=\delta_{kl}$. The measurement operators $M_{k}$ have the following explicit expressions,
\begin{equation}\label{GSIC-POVMs-construction}
M_{k}=\frac{1}{d^{2}}I_{d}+t\widetilde{F}_{k},
\end{equation}
where
\begin{equation*}
\widetilde{F}_{k}=
\begin{cases}
F-d(d+1)F_{k}, & k=1,\cdots,d^{2}-1,\\[2pt]
(d+1)F, & k=d^{2},
\end{cases}
\end{equation*}
with $F=\sum\limits_{k=1}^{d^{2}-1}F_{k}$. To ensure $M_{k}\geq0$, the parameter $t$ must satisfy
\begin{equation}\label{GSICt}
-\frac{1}{d^{2}}\frac{1}{\lambda_{\max}}\leq t \leq\frac{1}{d^{2}}\frac{1}{|\lambda_{\min}|},
\end{equation}
where $\lambda_{\max}$ and $\lambda_{\min}$ are the maximum and minimum eigenvalues, respectively, of all $\widetilde{F}_{k}$. Moreover, the parameter $a$ in Eq.~(\ref{GSICdef}) is given by
\begin{equation}\label{GSICa}
a=\frac{1}{d^{3}}+t^{2}(d-1)(d+1)^{3}.
\end{equation}

2) {\bf MUMs.} If $N=d+1$ and $M=d$, the $(N,M)$-POVMs reduce to MUMs \cite{KalevGour2014}. A set of POVMs $\{P^{(b)}\}_{b=1}^{d+1}$ on $\mathbb{C}^{d}$, where $P^{(b)}=\{P_{n}^{(b)}\}_{n=1}^{d}$, is called MUMs if
\begin{equation}\label{MUMs-def}
\begin{aligned}
\mathrm{Tr}\left(P^{(b)}_{n}\right)&=1,\\[2pt]
\mathrm{Tr}\left[\left(P^{(b)}_{n}\right)^{2}\right]&=\kappa,\\[2pt]
\mathrm{Tr}\left(P^{(b)}_{m}P^{(b)}_{n}\right)&=\frac{1-\kappa}{d-1}\quad(m \neq n),\\[2pt]
\mathrm{Tr}\left(P^{(a)}_{m}P^{(b)}_{n}\right)&=\frac{1}{d}\quad(a \neq b),
\end{aligned}
\end{equation}
where the parameter $\kappa$ satisfies $\frac{1}{d}<\kappa\leq1$. Let $\{F_{b,n}\mid b=1,\cdots,d+1;\,n=1,\cdots,d-1\}$ be $d^{2}-1$ Hermitian and traceless operators such that $\mathrm{Tr}(F_{a,m}F_{b,n})=\delta_{ab}\delta_{mn}$. The measurement operators of MUMs have the following explicit expressions,
\begin{equation}\label{MUMs-construction}
P_{n}^{(b)}=\frac{1}{d}I_{d}+t\widetilde{F}^{(b)}_{n},
\end{equation}
where
\begin{equation*}
\widetilde{F}^{(b)}_{n}=
\begin{cases}
F^{(b)}-\left(d+\sqrt{d}\right)F_{b,n}, & n=1,\cdots,d-1,\\[2pt]
\left(\sqrt{d}+1\right)F^{(b)}, & n=d,
\end{cases}
\end{equation*}
with $F^{(b)}=\sum\limits_{n=1}^{d-1}F_{b,n}$, $b=1,\cdots,d,d+1$. To ensure $P^{(b)}_{n}\geq0$, the parameter $t$ must satisfy
\begin{equation}\label{MUMs-t}
-\frac{1}{d}\frac{1}{\lambda_{\max}}\leq t \leq\frac{1}{d}\frac{1}{|\lambda_{\min}|},
\end{equation}
where $\lambda_{\max}=\max_{b}\lambda_{\max}^{(b)}$ and $\lambda_{\min}=\min_{b}\lambda_{\min}^{(b)}$, with $\lambda_{\max}^{(b)}$ ($\lambda_{\min}^{(b)}$) being the maximum (minimum) eigenvalue of all $\widetilde{F}^{(b)}_{n}$. The parameter $\kappa$ in Eq.~(\ref{MUMs-def}) is given by
\begin{equation}\label{MUMs-kappa}
\kappa=\frac{1}{d}+t^{2}(d-1)\left(\sqrt{d}+1\right)^{2}.
\end{equation}

In 2025, Lu \textit{et al.} \cite{LuPRA062424} constructed the matrix \(\mathcal{M}_{\alpha,\beta}^{l}(\rho_{AB})\) for bipartite states \(\rho_{AB}\) as follows:
\begin{align}\label{LocalMeasurementRealignment01}
\mathcal{M}_{\alpha,\beta}^{l}(\rho_{AB})=
\begin{pmatrix}
\alpha\beta J_{l \times l} & \alpha\omega_{l}(\sigma)^T \\
\beta \omega_{l}(\tau) & \mathcal{P}(\rho_{AB})
\end{pmatrix},
\end{align}
where
\begin{align}\label{SubsystemsMeanmentResults}
\tau = \left(
\begin{array}{c}
\mathrm{Tr}(E_{1,1}^A \rho_A) \\
\mathrm{Tr}(E_{1,2}^A \rho_A) \\
\vdots \\
\mathrm{Tr}(E_{1,M_A}^A \rho_A) \\
\mathrm{Tr}(E_{2,1}^A \rho_A) \\
\mathrm{Tr}(E_{2,2}^A \rho_A) \\
\vdots \\
\mathrm{Tr}(E_{2,M_A}^A \rho_A) \\
\vdots \\
\mathrm{Tr}(E_{N_A,M_A}^A \rho_A)
\end{array}
\right), \qquad
\sigma = \left(
\begin{array}{c}
\mathrm{Tr}(E^B_{1,1} \rho_B) \\
\mathrm{Tr}(E^B_{1,2} \rho_B) \\
\vdots \\
\mathrm{Tr}(E^B_{1,M_B} \rho_B) \\
\mathrm{Tr}(E^B_{2,1} \rho_B) \\
\mathrm{Tr}(E^B_{2,2} \rho_B) \\
\vdots \\
\mathrm{Tr}(E^B_{2,M_B} \rho_B) \\
\vdots \\
\mathrm{Tr}(E^B_{N_B,M_B} \rho_B)
\end{array}
\right),
\end{align}
with $\rho_{A}$ ($\rho_{B}$) being the reduced density matrix obtained by tracing over the subsystem $\mathcal{H}_B$ ($\mathcal{H}_A$) of $\rho_{AB}$, $\alpha$ and $\beta$ are real numbers, $l$ is a natural number, $J_{l \times l}$ is the matrix with all $l \times l$ entries being $1$, $\omega_{l}(X)=(\underbrace{X, \cdots, X}_{l\text{ columns}})$, and $T$ denotes the transpose of a matrix. They showed that for any separable state $\rho_{AB}$ on $\mathcal{H}_{A}\otimes\mathcal{H}_{B}$, the following inequality holds:
\begin{eqnarray}\label{Lu1}
\left\|\mathcal{M}_{\alpha, \beta}^l \left(\rho_{AB}\right)\right\|_{\mathrm{tr}} \leq \sqrt{\left(l \mu^{2}+\frac{(d_{A}-1)(M_{A}^2x_A+d_{A}^2)}{d_{A}M_{A}(M_{A}-1)}\right)\left(l \nu^{2}+\frac{(d_{B}-1)(M_{B}^2x_B+d_{B}^2)}{d_{B}M_{B}(M_{B}-1)}\right)}.
\end{eqnarray}

Subsequently, Lu \textit{et al}. \cite{LuAQTe70344} further generalized this framework and optimized this criterion under a more general parameterized structure. They constructed the matrix $\mathcal{Q}_{\bm{a}, \bm{b}}(\rho_{AB})$ for a bipartite state $\rho_{AB}$,
\begin{equation}\label{LocalMeasurementRealignment02}
\mathcal{Q}_{\boldsymbol{a}, \boldsymbol{b}}(\rho_{AB})=
\begin{pmatrix}
	\boldsymbol{a}\boldsymbol{b}^{T} & \boldsymbol{a} \sigma^{T}\\
	\tau \boldsymbol{b}^{T} & \mathcal{P}(\rho_{AB})
\end{pmatrix},
\end{equation}
where $\boldsymbol{a}\in\mathbb{R}^{m}$, $\boldsymbol{b}\in\mathbb{R}^{n}$, $\tau$ and $\sigma$  are defined in Eq.~(\ref{SubsystemsMeanmentResults}). They demonstrated that for any separable state \(\rho_{AB}\) on \(\mathcal{H}_{A}\otimes\mathcal{H}_{B}\), the following inequality holds:
\begin{equation}\label{Lu2}
\|\mathcal{Q}_{\mathbf{a}, \bm{b}}(\rho_{AB})\|_{\mathrm{tr}}\leq\sqrt{\left(|\boldsymbol{a}|^{2}+\frac{(d_{A}-1)(M_{A}^{2}x_{A}+d_{A}^{2})}{d_{A}M_{A}(M_{A}-1)}\right)\left(|\boldsymbol{b}|^{2}+\frac{(d_{B}-1)(M_{B}^{2}x_{B}+d_{B}^{2})}{d_{B}M_{B}(M_{B}-1)}\right)}.
\end{equation}

\section{Entanglement detection based on $(N,M)$-POVMs}\label{sec3}

We note that, in the above two constructions (\ref{LocalMeasurementRealignment01}) and  (\ref{LocalMeasurementRealignment02}), the upper-left matrix is always of rank one, imposing certain limitations on the amount of information the construction can encode. By relaxing this inherent rank-one constraint, we introduce a more general correlation matrix denoted by \(\mathcal{Q}_{P,Q,\mu,\nu}(\rho_{AB})\). This matrix preserves the advantages of the original construction while enhancing its capability of detecting different families of entangled states. In what follows, we elaborate on the construction of this new criterion and its theoretical underpinnings.

We construct the matrix $\mathcal{Q}_{P,Q,\mu,\nu}(\rho_{AB})$ for a bipartite state $\rho_{AB}$ as follows,
\begin{equation}\label{LocalMeasurementRealignment}
\mathcal{Q}_{P,Q,\mu,\nu}(\rho_{AB})=
\begin{pmatrix}
PQ^T & P\mu\sigma^T \\[2pt]
\tau\nu^TQ^T & \mathcal{P}(\rho_{AB})
\end{pmatrix},
\end{equation}
where $P\in\mathbb{R}^{m \times r}, Q\in\mathbb{R}^{n \times r}$, and $\mu, \nu\in\mathbb{R}^{r \times 1}$ with $\nu^T\mu=1$. In this case, one may choose appropriate matrices $P$ and $Q$ such that the rank of \(PQ^T\) can be any positive integer. For instance, if \(m\geq r\), \(n\geq r\), and \(\mathrm{rank}(P)=\mathrm{rank}(Q)=r\), then \(\mathrm{rank}(PQ^T)=r\). If $r=1$, $\mu=\nu=1$, $m=n=l$, and $P=(\underbrace{\alpha,\cdots,\alpha}_{l\text{ columns}})^T$, $Q=(\underbrace{\beta,\cdots,\beta}_{l\text{ columns}})^T$, the matrix (\ref{LocalMeasurementRealignment}) reduces to the matrix (\ref{LocalMeasurementRealignment01}). If $r=1$, $\mu=\nu=1$, $P=\bm{a}$, and $Q=\bm{b}$, then the matrix (\ref{LocalMeasurementRealignment}) reduces to the matrix (\ref{LocalMeasurementRealignment02}).

Denote
\begin {align*}
\mathfrak {M}(P,Q,\mu,\nu,M_A,&M_B,d_A,d_B,x_A,x_B) = \\
&\left\|\left (P^TP+\frac {(d_{A}-1)(M_{A}^{2} x_{A}+d_{A}^{2})}{d_{A} M_{A}(M_{A}-1)}\nu\nu^T\right)^{1/2}\left (Q^TQ+\frac {(d_{B}-1)(M_{B}^{2} x_{B}+d_{B}^{2})}{d_{B} M_{B}(M_{B}-1)}\mu\mu^T\right)^{1/2}\right\|_{\mathrm {tr}}.
\end {align*}
We then arrive at the following separability criterion:
\begin{theorem}\label{thm1}
If \(\rho_{AB}\) is a bipartite separable state, then
\begin{equation}\label{SeparabilityUpperBound}
\|\mathcal{Q}_{P,Q,\mu,\nu}(\rho_{AB})\|_{\mathrm{tr}}\leq\mathfrak{M}(P,Q,\mu,\nu,M_A,M_B,d_A,d_B,x_A,x_B),
\end{equation}
where \(\mathcal{Q}_{P,Q,\mu,\nu}(\rho_{AB})\) is defined in (\ref{LocalMeasurementRealignment}).
\end{theorem}

\begin{proof}
If $\rho_{AB}=\rho_A\otimes\rho_B$ is a product state, we have
\begin{align*}
\|\mathcal{Q}_{P,Q,\mu,\nu}\left(\rho_{A} \otimes \rho_{B}\right)\|_{\mathrm{tr}}
&=\left\|\left(
\begin{array}{cc}
PQ^{T} & P\mu\sigma^T \\
\tau \nu^T Q^T & \tau\sigma^T
\end{array}
\right)\right\|_{\mathrm{tr}} \\
&=\left\|\left(
\begin{array}{c}
P \\
\tau\nu^T
\end{array}
\right)
\left(
\begin{array}{cc}
Q^T & \mu\sigma^T
\end{array}
\right)
\right\|_{\mathrm{tr}} \\
&=\left\Vert\Big(P^TP+\tau^T\tau\nu\nu^T\Big)^{1/2}\Big(Q^TQ+\sigma^T\sigma\mu\mu^T\Big)^{1/2}\right\Vert_{\mathrm{tr}} \\
&\leq\mathfrak{M}(P,Q,\mu,\nu,M_A,M_B,d_A,d_B,x_A,x_B),
\end{align*}
where $\|XY^T\|_{\mathrm{tr}}=\left\|\left(X^TX\right)^{1/2}\left(Y^TY\right)^{1/2}\right\|_{\mathrm{tr}}$.
Hence for any separable state $\rho_{AB}$ of the form $\rho_{AB}=\sum\limits_k p_k \, \rho^A_k\otimes\rho^B_k$, we have
\begin{equation*}
\begin{split}
\|\mathcal{Q}_{P,Q,\mu,\nu}\left(\rho_{AB}\right)\|_{\mathrm{tr}}
&=\left\|\sum\limits_{k}p_{k}\mathcal{Q}_{P,Q,\mu,\nu}\left(\rho^{A}_{k} \otimes \rho^{B}_{k}\right)\right\|_{\mathrm{tr}} \\
&\leq \sum\limits_{k}p_{k}\Big\|\mathcal{Q}_{P,Q,\mu,\nu}\Big(\rho^{A}_{k} \otimes \rho^{B}_{k}\Big)\Big\|_{\mathrm{tr}}\\
&\leq\sum_kp_k\mathfrak{M}(P,Q,\mu,\nu,M_A,M_B,d_A,d_B,x_A,x_B) \\
&=\mathfrak{M}(P,Q,\mu,\nu,M_A,M_B,d_A,d_B,x_A,x_B),
\end{split}
\end{equation*}
where the first equality follows from
\begin{align*}
\mathcal{Q}_{P,Q,\mu,\nu}\left(\sum\limits_{k}p_{k}\rho^A_{k} \otimes \rho^{B}_{k} \right)&=
\begin{pmatrix}
	PQ^{T} & P\mu\sigma^{T}\\
	\tau \nu^{T} Q^T & \mathcal{P}(\sum\limits_{k}p_{k}\rho^{A}_{k} \otimes \rho^{B}_{k})
\end{pmatrix} \\
&=\begin{pmatrix}
	\sum\limits_{k}p_{k}PQ^{T} & \sum\limits_{k}p_{k}P\mu\sigma_{k}^{T}\\
	\sum\limits_{k}p_{k}\tau_{k} \nu^{T} Q^T & \sum\limits_{k}p_{k}\mathcal{P}(\rho^{A}_{k} \otimes \rho^{B}_{k})
\end{pmatrix}\\
&=\sum\limits_{k}p_{k}\begin{pmatrix}
	PQ^{T} & P \mu \sigma_{k}^{T}\\
	\tau_{k} \nu^{T} Q^T & \mathcal{P}(\rho^{A}_{k} \otimes \rho^{B}_{k})
\end{pmatrix} \\
&=\sum\limits_{k}p_{k}\mathcal{Q}_{P,Q,\mu,\nu}\left(\rho^{A}_{k}\otimes \rho^{B}_{k}\right),
\end{align*}
with
\begin{align*}
	\tau = \left(\begin{array}{c}
		\mathrm{Tr}(E_{1,1} \sum\limits_{k}p_{k}\rho^{A}_{k}) \\
		\mathrm{Tr}(E_{1,2} \sum\limits_{k}p_{k}\rho^{A}_{k}) \\
		\vdots \\
		\mathrm{Tr}(E_{N_A,M_A} \sum\limits_{k}p_{k}\rho^{A}_{k})
	\end{array}\right), \qquad
		\sigma = \left(\begin{array}{c}
		\mathrm{Tr}(E_{1,1} \sum\limits_{k}p_{k}\rho^{B}_{k}) \\
		\mathrm{Tr}(E_{1,2} \sum\limits_{k}p_{k}\rho^{B}_{k}) \\
		\vdots \\
		\mathrm{Tr}(E_{N_B,M_B} \sum\limits_{k}p_{k}\rho^{B}_{k})
	\end{array}\right),
\end{align*}
\begin{align*}
		\tau_k = \left(\begin{array}{c}
		\mathrm{Tr}(E_{1,1} \rho^{A}_{k}) \\
		\mathrm{Tr}(E_{1,2} \rho^{A}_{k}) \\
		\vdots \\
		\mathrm{Tr}(E_{N_A,M_A} \rho^{A}_{k})
	\end{array}\right), \qquad
		\sigma_k = \left(\begin{array}{c}
		\mathrm{Tr}(E_{1,1} \rho^{B}_{k}) \\
		\mathrm{Tr}(E_{1,2} \rho^{B}_{k}) \\
		\vdots \\
		\mathrm{Tr}(E_{N_B,M_B} \rho^{B}_{k})
	\end{array}\right).
\end{align*}
Thus, the inequality (\ref{SeparabilityUpperBound}) holds.
\end{proof}

In particular, if we choose the \((N,M)\)-POVMs to be GSIC-POVMs and MUMs, we obtain the following corollaries, respectively.

\begin{corollary}\label{cor1}
Consider two GSIC-POVMs $\{M^A_{\alpha}\}_{\alpha=1}^{d_A^{2}}$ with parameter $a_A$ and $\{M^B_{\beta}\}_{\beta=1}^{d_B^{2}}$ with parameter $a_B$.
If a state $\rho_{AB}$ on $\mathcal{H}_{A} \otimes \mathcal{H}_{B}$ is separable, then
\begin{eqnarray}\label{Cor1SeparabilityUpperBound}
\left\|\mathcal{Q}_{P,Q,\mu,\nu}\left(\rho_{AB}\right)\right\|_{\mathrm{tr}} \leq \norm{\left(P^TP+\frac{a_{A}d_{A}^{2}+1}{d_{A}(d_{A}+1)}\nu\nu^T\right)^{1/2}\left(Q^TQ+\frac{a_{B}d_{B}^{2}+1}{d_{B}(d_{B}+1)}\mu\mu^T\right)^{1/2}}_{\mathrm{tr}},
\end{eqnarray}
where
\begin{align*}
\mathcal{Q}_{P,Q,\mu,\nu}(\rho_{AB})
=\begin{pmatrix}
	P Q^T & P\mu\xi^T \\
	\zeta \nu^T Q^T & \mathcal{G}(\rho_{AB})
\end{pmatrix},
\end{align*}
with
\begin{equation*}
\zeta = \left(\begin{array}{c}
\mathrm{Tr}(M_1^A \rho_A) \\
\mathrm{Tr}(M_2^A \rho_A) \\
\vdots \\
\mathrm{Tr}(M_{d^2_A}^A \rho_A)
\end{array}\right), \qquad
\xi = \left(\begin{array}{c}
\mathrm{Tr}(M_1^B \rho_B) \\
\mathrm{Tr}(M_2^B \rho_B) \\
\vdots \\
\mathrm{Tr}(M_{d^2_B}^B \rho_B)
\end{array}\right),
\end{equation*}
where $\mathcal{G}(\rho_{AB})$ is a matrix with entries given by $\mathrm{Tr}[(M^A_{\alpha}\otimes M^B_{\beta})\rho_{AB}]$.
\end{corollary}

\begin{corollary}\label{cor2}
Let $\{P^{(b_A)}\}_{b_A=1}^{d_A+1}$ ($\{Q^{(b_B)}\}_{b_B=1}^{d_B+1}$) be the MUMs on the subsystem $\mathcal{H}_A$ ($\mathcal{H}_B$) with the parameter $\kappa_A$ ($\kappa_B$), where $P^{(b_A)}=\{P^{(b_A)}_{n_A}\}_{n_A=1}^{d_A}$ ($Q^{(b_B)}=\{Q^{(b_B)}_{n_B}\}_{n_B=1}^{d_B}$). Denote by $\mathcal{J}(\rho_{AB})$ the matrix with entries given by the probabilities $\mathrm{Tr}[(P^{(b_A)}_{n_A} \otimes Q^{(b_B)}_{n_B})\rho_{AB}]$, where $b_A, n_A$ and $b_B, n_B$ are the row and the column indices, respectively. For any separable state $\rho_{AB}$ on $\mathcal{H}_{A} \otimes \mathcal{H}_B$, we have
\begin{eqnarray}\label{Cor2SeparabilityUpperBound}
\left\|\mathcal{Q}_{P,Q,\mu,\nu}(\rho_{AB})\right\|_{\mathrm{tr}} \leq \norm{\Big(P^TP + (1+\kappa_A)\nu\nu^T\Big)^{1/2}\Big(Q^TQ + (1+\kappa_B)\mu\mu^T\Big)^{1/2}}_{\mathrm{tr}},
\end{eqnarray}
where
\begin{align*}
\mathcal{Q}_{P,Q,\mu,\nu}(\rho_{AB})=
\begin{pmatrix}
	P Q^T & P\mu\eta^T \\
	\theta \nu^T Q^T & \mathcal{J}(\rho_{AB})
\end{pmatrix},
\end{align*}
with
\begin{eqnarray*}
\theta = \left(\begin{array}{c}
\mathrm{Tr}(P^{(1)}_1 \rho_A) \\
\mathrm{Tr}(P^{(1)}_2 \rho_A) \\
\vdots \\
\mathrm{Tr}(P^{(1)}_{d_A} \rho_A)\\
\mathrm{Tr}(P^{(2)}_1 \rho_A) \\
\mathrm{Tr}(P^{(2)}_2 \rho_A) \\
\vdots \\
\mathrm{Tr}(P^{(2)}_{d_A} \rho_A)\\
\vdots \\
\mathrm{Tr}(P^{(d_A+1)}_{d_A} \rho_A)
\end{array}\right), \qquad
\eta = \left(\begin{array}{c}
\mathrm{Tr}(Q^{(1)}_1 \rho_B) \\
\mathrm{Tr}(Q^{(1)}_2 \rho_B) \\
\vdots \\
\mathrm{Tr}(Q^{(1)}_{d_B} \rho_B)\\
\mathrm{Tr}(Q^{(2)}_1 \rho_B) \\
\mathrm{Tr}(Q^{(2)}_2 \rho_B) \\
\vdots \\
\mathrm{Tr}(Q^{(2)}_{d_B} \rho_B)\\
\vdots \\
\mathrm{Tr}(Q^{(d_B+1)}_{d_B} \rho_B)
\end{array}\right).
\end{eqnarray*}
\end{corollary}

We illustrate our results through several examples.

\begin{example}\label{eg1}
We consider the following state obtained by mixing $\rho_{AB}$ with white noise,
\begin{equation*}
\rho_{p}=\frac{1-p}{9}I_{9}+p\rho_{AB},
\end{equation*}
where
\begin{equation*}
\rho_{AB} = \frac{1}{4}\left(I_9-\sum\limits_{k=0}^{4}\ket{\phi_{k}}\bra{\phi_{k}}\right)
\end{equation*}
is a $3\times 3$ PPT entangled state with
\begin{align*}
&\ket{\phi_{0}}=\ket{0}\otimes\frac{1}{\sqrt{2}}(\ket{0}-\ket{1}),~~ \ket{\phi_{1}}=\frac{1}{\sqrt{2}}(\ket{0}-\ket{1})\otimes\ket{2}, \\
&\ket{\phi_{2}}=\ket{2}\otimes\frac{1}{\sqrt{2}}(\ket{1}-\ket{2}),~~  \ket{\phi_{3}}=\frac{1}{\sqrt{2}}(\ket{1}-\ket{2})\otimes\ket{0},\\
&\ket{\phi_{4}}=\frac{1}{\sqrt{3}}(\ket{0}+\ket{1}+\ket{2})\otimes\dfrac{1}{\sqrt{3}}(\ket{0}+\ket{1}+\ket{2}).
\end{align*}

Consider the $(8,2)$-POVMs with its Hermitian basis operators $G_{\alpha, k}$ defined via the Gell-Mann matrices,
\begin{align*}
	& G_{11}=\dfrac{1}{\sqrt{2}}
	\begin{pmatrix}
		0 & 1 & 0\\
		1 & 0 & 0\\
		0 & 0 & 0
	\end{pmatrix},~~
	G_{21}=\dfrac{1}{\sqrt{2}}
	\begin{pmatrix}
		0 & -\mathrm{i} & 0\\
		\mathrm{i} & 0 & 0\\
		0 & 0 & 0
	\end{pmatrix},~~
	G_{31}=\dfrac{1}{\sqrt{2}}
	\begin{pmatrix}
		0 & 0 & 1\\
		0 & 0 & 0\\
		1 & 0 & 0
	\end{pmatrix},~~
	G_{41}=\dfrac{1}{\sqrt{2}}
	\begin{pmatrix}
		0 & 0 & -\mathrm{i}\\
		0 & 0 & 0\\
		\mathrm{i} & 0 & 0
	\end{pmatrix}, \\
	& G_{51}=\dfrac{1}{\sqrt{2}}
	\begin{pmatrix}
		0 & 0 & 0\\
		0 & 0 & 1\\
		0 & 1 & 0
	\end{pmatrix},~~
	G_{61}=\dfrac{1}{\sqrt{2}}
	\begin{pmatrix}
		0 & 0 & 0\\
		0 & 0 & -\mathrm{i}\\
		0 & \mathrm{i} & 0
	\end{pmatrix},~~
	G_{71}=\dfrac{1}{\sqrt{2}}
	\begin{pmatrix}
		1 & 0 & 0\\
		0 & -1 & 0\\
		0 & 0 & 0
	\end{pmatrix},~~
	G_{81}=\dfrac{1}{\sqrt{6}}
	\begin{pmatrix}
		1 & 0 & 0\\
		0 & 1 & 0\\
		0 & 0 & -2
	\end{pmatrix}.
\end{align*}
Here, the parameter $x=\frac{3}{4}+(\sqrt{2}+1)^{2}t^{2}$ with $t\in[-0.2536,0.2536]$. We take
\begin{equation*}
	P=
	\begin{pmatrix}
		0.03 & 0.05 \\
		0.14 & 0.08 \\
		0.13 & 0.11 \\
		0.04 & 0.08
	\end{pmatrix},\quad
	Q=
	\begin{pmatrix}
		0.05 & 0.07 \\
		0.13 & 0.07 \\
		0.14 & 0.12
	\end{pmatrix},\quad
	\mu=\left(
	\begin{array}{c}
		\frac{1}{\sqrt{2}} \\[3pt]
		\frac{1}{\sqrt{2}}
	\end{array}\right),\quad
	\nu=\left(
	\begin{array}{c}
		\frac{1}{\sqrt{2}} \\[3pt]
		\frac{1}{\sqrt{2}}
	\end{array}\right),
\end{equation*}
and $t = 0.01$ in (\ref{LocalMeasurementRealignment}).

Denote
\begin{equation*}
f_1(p) = \left\|\mathcal{Q}_{P, Q, \mu, \nu}\left(\rho_{p}\right)\right\|_{\mathrm{tr}} -\mathfrak{M}(P,Q,\mu,\nu,d_A,d_B,M_A,M_B,x_A,x_B).
\end{equation*}
Using Theorem \ref{thm1}, we obtain that $\rho_p$ is entangled when $f_1(p)$ is positive, i.e., when $0.88217822 \leq p \leq 1$.

Concerning the Theorem $1$ in Ref.~\cite{LuAQTe70344}, we denote
\begin{equation*}
g_1(p) = \left\|\mathcal{Q}_{\boldsymbol{a},\boldsymbol{b}}(\rho_{p})\right\|_{\mathrm{tr}} - \sqrt{\left(|\boldsymbol{a}|^{2}+\frac{(d_{A}-1)(M_{A}^{2}x_{A}+d_{A}^{2})}
{d_{A}M_{A}(M_{A}-1)}\right)\left(|\boldsymbol{b}|^{2}+\frac{(d_{B}-1)(M_{B}^{2}x_{B}
+d_{B}^{2})}{d_{B}M_{B}(M_{B}-1)}\right)}.
\end{equation*}
Take $\bm{a} = (0.1, 0.1)^T$, $\bm{b} = (0.05, 0.07)^T$ and $t=0.01$, from (\ref{Lu2}) we obtain that
$\rho_p$ is entangled when $g_1(p)$ is positive, i.e., when $0.88217901 \leq p \leq 1$.

With respect to the Theorem $1$ in Ref.~\cite{LuPRA062424}, we denote
\begin{equation*}
h_1(p) = \left\|\mathcal{M}_{\alpha, \beta}^l \left(\rho_{p}\right)\right\|_{\mathrm{tr}} - \sqrt{\left(l \alpha^{2}+\frac{(d_{A}-1)(M_{A}^2x_A+d_{A}^2)}{d_{A}M_{A}(M_{A}-1)}\right)\left(l \beta^{2}+\frac{(d_{B}-1)(M_{B}^2x_B+d_{B}^2)}{d_{B}M_{B}(M_{B}-1)}\right)}.
\end{equation*}
Take $\alpha = 0.1$, $\beta = 0.05$, $l = 2$ and $t = 0.01$. 
From (\ref{Lu1}) we obtain that $\rho_p$ is entangled when $h_1(p)$ is positive, i.e., when $0.88217942 \leq p \leq 1$.

The graphs of \(f_1(p)\), \(g_1(p)\) and \(h_1(p)\) are plotted in FIG. \ref{example1fig1}. Obviously our method detects a broader range of entangled states compared with the approaches in Refs. \cite{LuPRA062424,LuAQTe70344}.
\begin{figure}[htbp]
   \includegraphics[width=0.8\textwidth]{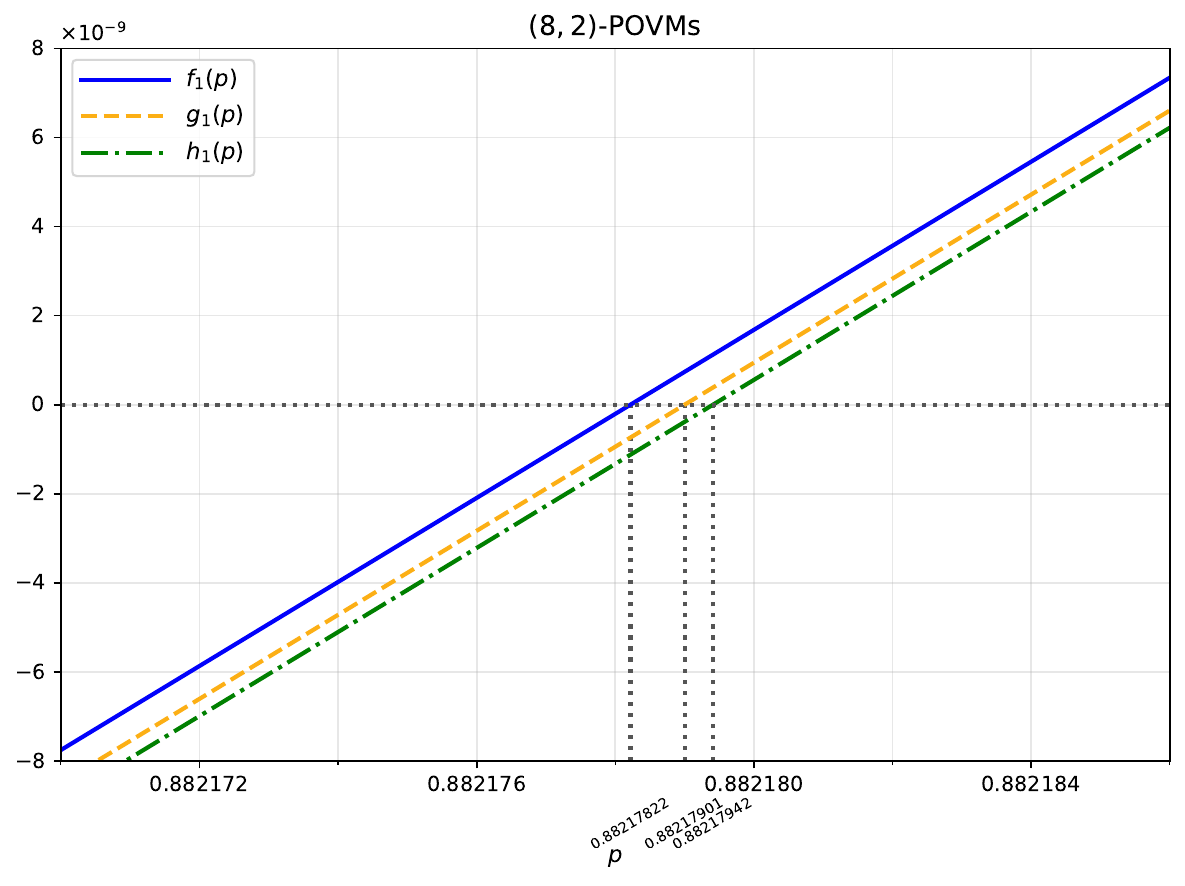}
    \caption{$f_1(p)$ derived from Theorem \ref{thm1} (solid blue line). $g_1(p)$ from Theorem 1 in Ref. \cite{LuAQTe70344} (dashed orange line). $h_1(p)$ presented in Theorem 1 in Ref. \cite{LuPRA062424} (dot-dashed green line).}
    \label{example1fig1}
\end{figure}

When $N = 1$ and $M = 9$, Theorem \ref{thm1} reduces to Corollary \ref{cor1}. In this case, the Hermitian basis operators $G_{\alpha, k}$ are given by the corresponding Gell-Mann matrices:
\begin{align*}
	&G_{11}=\frac{1}{\sqrt{2}}\left(
	\begin{array}{ccc}
		1 & 0 & 0 \\
		0 & -1 & 0 \\
		0 & 0 & 0
	\end{array}
	\right),~~
	G_{12}=\frac{1}{\sqrt{2}}\left(
	\begin{array}{ccc}
		0 & 1 & 0 \\
		1 & 0 & 0 \\
		0 & 0 & 0
	\end{array}
	\right),~~
	G_{13}=\frac{1}{\sqrt{2}}\left(
	\begin{array}{ccc}
		0 & 0 &1 \\
		0 & 0 & 0 \\
		1 & 0 & 0
	\end{array}
	\right),~~
	G_{14}=\frac{1}{\sqrt{2}}\left(
	\begin{array}{ccc}
		0 & -\mathrm{i} & 0 \\
		\mathrm{i} & 0 & 0 \\
		0 & 0 & 0
	\end{array}
	\right),~~\\
	&G_{15}=\frac{1}{\sqrt{6}}\left(
	\begin{array}{ccc}
		1 & 0 & 0 \\
		0 & 1 & 0 \\
		0 & 0 & -2
	\end{array}
	\right),~~
	G_{16}=\frac{1}{\sqrt{2}}\left(
	\begin{array}{ccc}
		0 & 0 & 0 \\
		0 & 0 & 1 \\
		0 & 1 & 0
	\end{array}
	\right),~~
	G_{17}=\frac{1}{\sqrt{2}}\left(
	\begin{array}{ccc}
		0 & 0 & -\mathrm{i} \\
		0 & 0 & 0 \\
		\mathrm{i} & 0 & 0
	\end{array}
	\right),~~
	G_{18}=\frac{1}{\sqrt{2}}\left(
	\begin{array}{ccc}
		 0 & 0 & 0 \\
		 0 & 0 & -\mathrm{i} \\
		 0 & \mathrm{i} & 0
	\end{array}
	\right).
\end{align*}
Here, the corresponding parameter is $a=\frac{1}{27}+128t^{2}$ with $t\in[-0.012, 0.012]$.

To apply Corollary \ref{cor1}, we take
\begin{equation*}
	P=
	\begin{pmatrix}
		0.08 & 0.07 \\
		0.05 & -0.02 \\
		0.13 & 0.12 \\
		-0.02 & 0.12
	\end{pmatrix},\quad
	Q=
	\begin{pmatrix}
		0.12 & 0.1 \\
		0.08 & 0.06 \\
		0.03 & 0.12
	\end{pmatrix},\quad
	\mu=
	\begin{pmatrix}
		\frac{1}{\sqrt{2}} \\[3pt]
		\frac{1}{\sqrt{2}}
	\end{pmatrix},\quad
	\nu=
	\begin{pmatrix}
		\frac{1}{\sqrt{2}} \\[3pt]
		\frac{1}{\sqrt{2}}
	\end{pmatrix},
\end{equation*}
and $t=0.01$. Denote
\begin{equation*}
f_2(p)=\left\|\mathcal{Q}_{P, Q, \mu, \nu}\left(\rho_{p}\right)\right\|_{\mathrm{tr}}-\norm{\left(P^TP+\frac{a_{A}d_{A}^{2}+1}
{d_{A}(d_{A}+1)}\nu\nu^T\right)^{1/2}\left(Q^TQ+\frac{a_{B}d_{B}^{2}+1}{d_{B}(d_{B}+1)}
\mu\mu^T\right)^{1/2}}_{\mathrm{tr}}.
\end{equation*}
Direct computation gives $f_2(p)>0$ when $0.882505676 \leq p \leq 1$, namely, $\rho_p$ is entangled for $0.882505676 \leq p \leq 1$.

Denote
\begin{equation*}
g_2(p)=\left\|\mathcal{Q}_{\boldsymbol{a}, \boldsymbol{b}} \left(\rho_p\right)\right\|_{\mathrm{tr}} -\sqrt{\left(|\boldsymbol{a}|^{2}+\frac{a_{A}d_{A}^{2}+1}{d_{A}(d_{A}+1)}\right)\left(|\boldsymbol{b}|^2
+\frac{a_{B}d_{B}^{2}+1}{d_{B}(d_{B}+1)}\right)}.
\end{equation*}
Taking $\boldsymbol{a}=(0.1, 0.1)^T$, $\boldsymbol{b} =(0.05, 0.07)^T$ and $t = 0.01$, one has
by Corollary 1 of Ref.~\cite{LuAQTe70344} that $\rho_p$ is entangled ($g_2(p)>0$) within the range $0.883472971 \leq p \leq 1$.

For the Corollary $1$ in Ref.~\cite{LuPRA062424}, define
\begin{equation*}
h_2(p)=\left\|\mathcal{M}_{\alpha, \beta}^l \left(\rho_p\right)\right\|_{\mathrm{tr}} - \sqrt{\left(l \alpha^{2}+\frac{a_{A}d_{A}^{2}+1}{d_{A}(d_{A}+1)}\right)\left(l \beta^{2}+\frac{a_{B}d_{B}^{2}+1}{d_{B}(d_{B}+1)}\right)}.
\end{equation*}
Taking $\alpha = 0.1$, $\beta = 0.05$, $l = 2$ and $t = 0.01$, one has that  $\rho_p$ is entangled ($h_2(p)>0$) for $0.883894443 \leq p \leq 1$.

Clearly, our result outperforms those of Refs.~\cite{LuPRA062424,LuAQTe70344}. FIG. \ref{example1fig2} displays $f_2(p)$, $g_2(p)$ and $h_2(p)$.
\begin{figure}[htbp]
   \includegraphics[width=0.8\textwidth]{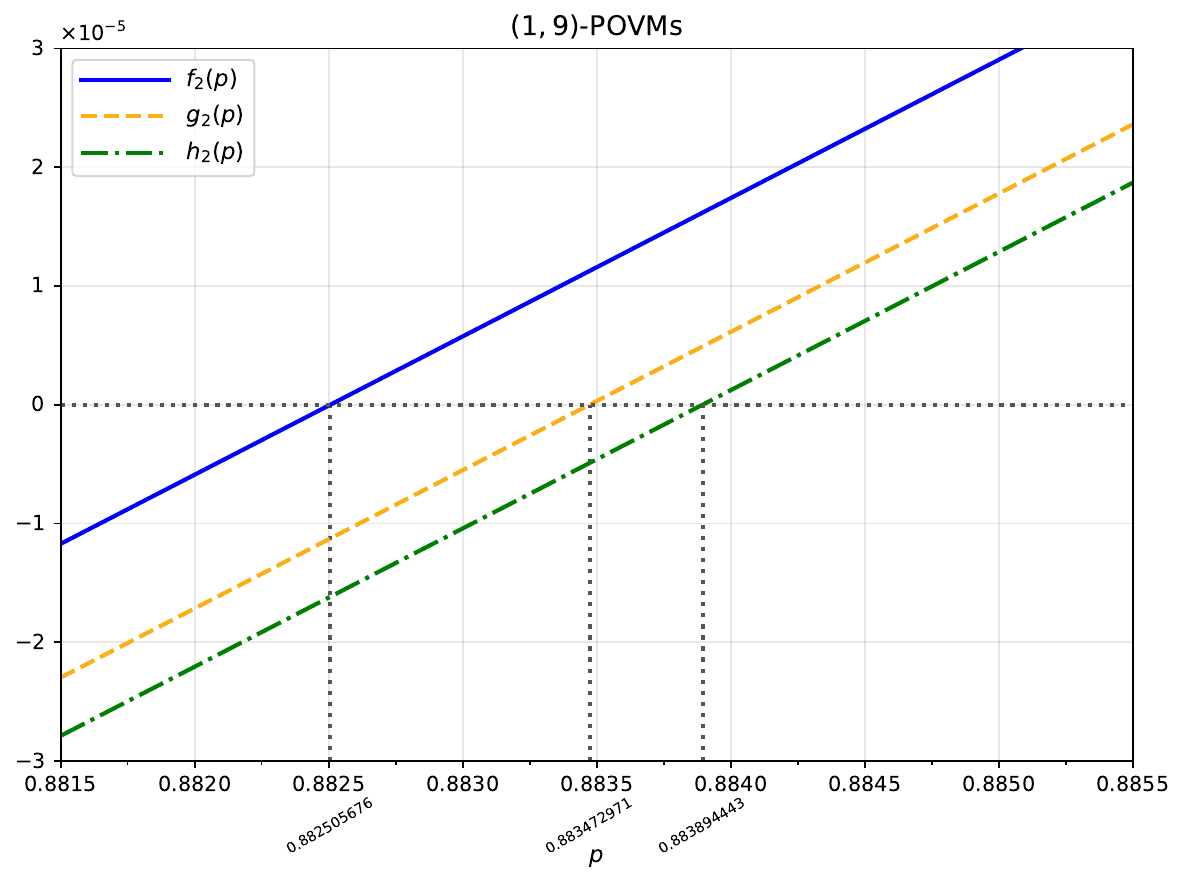}
    \caption{$f_2(p)$ derived from Corollary \ref{cor1} (solid blue line). $g_2(p)$ from Corollary 1 in Ref. \cite{LuAQTe70344} (dashed orange line). $h_2(p)$ presented in Corollary 1 in Ref. \cite{LuPRA062424} (dot-dashed green line).}
    \label{example1fig2}
\end{figure}

When $N = 4$ and $M = 3$, Theorem \ref{thm1} reduces to Corollary \ref{cor2}. In this case, the Hermitian basis operators $G_{\alpha, k}$ are given by the corresponding Gell-Mann matrices:
\begin{align*}
G_{11} &= \frac{1}{\sqrt{2}}
\left(
\begin{array}{ccc}
0 & 1 & 0 \\
1 & 0 & 0 \\
0 & 0 & 0
\end{array}
\right),~~
G_{12} = \frac{1}{\sqrt{2}}
\left(
\begin{array}{ccc}
0 & -\mathrm{i} & 0 \\
\mathrm{i} & 0 & 0 \\
0 & 0 & 0
\end{array}
\right),~~
G_{21} = \frac{1}{\sqrt{2}}
\left(
\begin{array}{ccc}
0 & 0 & 1 \\
0 & 0 & 0 \\
1 & 0 & 0
\end{array}
\right),~~
G_{22} = \frac{1}{\sqrt{2}}
\left(
\begin{array}{ccc}
0 & 0 & -\mathrm{i} \\
0 & 0 & 0 \\
\mathrm{i} & 0 & 0
\end{array}
\right),\\
G_{31} &= \frac{1}{\sqrt{2}}
\left(
\begin{array}{ccc}
0 & 0 & 0 \\
0 & 0 & 1 \\
0 & 1 & 0
\end{array}
\right),~~
G_{32} = \frac{1}{\sqrt{2}}
\left(
\begin{array}{ccc}
0 & 0 & 0 \\
0 & 0 & -\mathrm{i} \\
0 & \mathrm{i} & 0
\end{array}
\right),~~
G_{41} = \frac{1}{\sqrt{2}}
\left(
\begin{array}{ccc}
1 & 0 & 0 \\
0 & -1 & 0 \\
0 & 0 & 0
\end{array}
\right),~~
G_{42} = \frac{1}{\sqrt{6}}
\left(
\begin{array}{ccc}
1 & 0 & 0 \\
0 & 1 & 0 \\
0 & 0 & -2
\end{array}
\right).
\end{align*}
Here, the parameter $\kappa=\frac{1}{3}+2(\sqrt{3}+1)^2t^{2}$ with $t\in[-0.1092,0.1219]$.
We define
\begin{equation*}
f_3(p)=\left\|\mathcal{Q}_{P, Q, \mu, \nu}\left(\rho_{p}\right)\right\|_{\mathrm{tr}} - \norm{\Big(P^TP + (1+\kappa_A)\nu\nu^T\Big)^{1/2}\Big(Q^TQ + (1+\kappa_B)\mu\mu^T\Big)^{1/2}}_{\mathrm{tr}}.
\end{equation*}
In particular, for
\begin{equation*}
P=
\begin{pmatrix}
0.06 & 0.16 \\
0.14 & -0.02 \\
0.08 & 0.19 \\
0.08 & 0.09
\end{pmatrix},\quad
Q=
\begin{pmatrix}
0.11 & 0.14 \\
0.1 & 0.17 \\
0.11 & 0.07
\end{pmatrix},\quad
\mu=
	\begin{pmatrix}
		\frac{1}{\sqrt{2}} \\[3pt]
		\frac{1}{\sqrt{2}}
	\end{pmatrix},\quad
	\nu=
	\begin{pmatrix}
		\frac{1}{\sqrt{2}} \\[3pt]
		\frac{1}{\sqrt{2}}
	\end{pmatrix},
\end{equation*}
by using Corollary \ref{cor2} we have that $f_3(p)>0$ ($\rho_p$ is entangled) when $0.882186405 \leq p \leq 1$.

For $\boldsymbol{a} = (0.1, 0.1)^T$, $\boldsymbol{b} =(0.05, 0.07)^T$ and $t = 0.01$, Corollary 2 in Ref.~\cite{LuAQTe70344} gives rise to
\begin{equation*}
g_3(p)=\left\|\mathcal{Q}_{\boldsymbol{a},\boldsymbol{b}}(\rho_p)\right\|_{\mathrm{tr}} - \sqrt{\left(|\boldsymbol{a}|^2 + 1 + \kappa_A\right)\left(|\boldsymbol{b}|^2 + 1 + \kappa_B\right)}>0
\end{equation*}
for entangled $\rho_p$, i.e., in the range $0.882190974 \leq p \leq 1$. 

From Corollary $2$ in Ref.\cite{LuPRA062424}, for $\alpha = 0.1$, $\beta=0.05$, $l = 2$ and $t = 0.01$,  one has
\begin{align*}
h_3(p)=\left\|\mathcal{M}_{\alpha, \beta}^l \left(\rho_p\right)\right\|_{\mathrm{tr}} - \sqrt{\left(l\alpha^{2} + 1 + \kappa_A\right)\left(l\beta^{2} + 1 + \kappa_B\right)}>0,
\end{align*}
if $\rho_p$ is entangled, which gives rise to $0.882194618 \leq p \leq 1$. A comparison with the results in Refs. \cite{LuPRA062424,LuAQTe70344}, illustrated in FIG.~\ref{example1fig3}, shows that our method detects a wider range of entangled states.
\begin{figure}
	\includegraphics[width=0.80\textwidth]{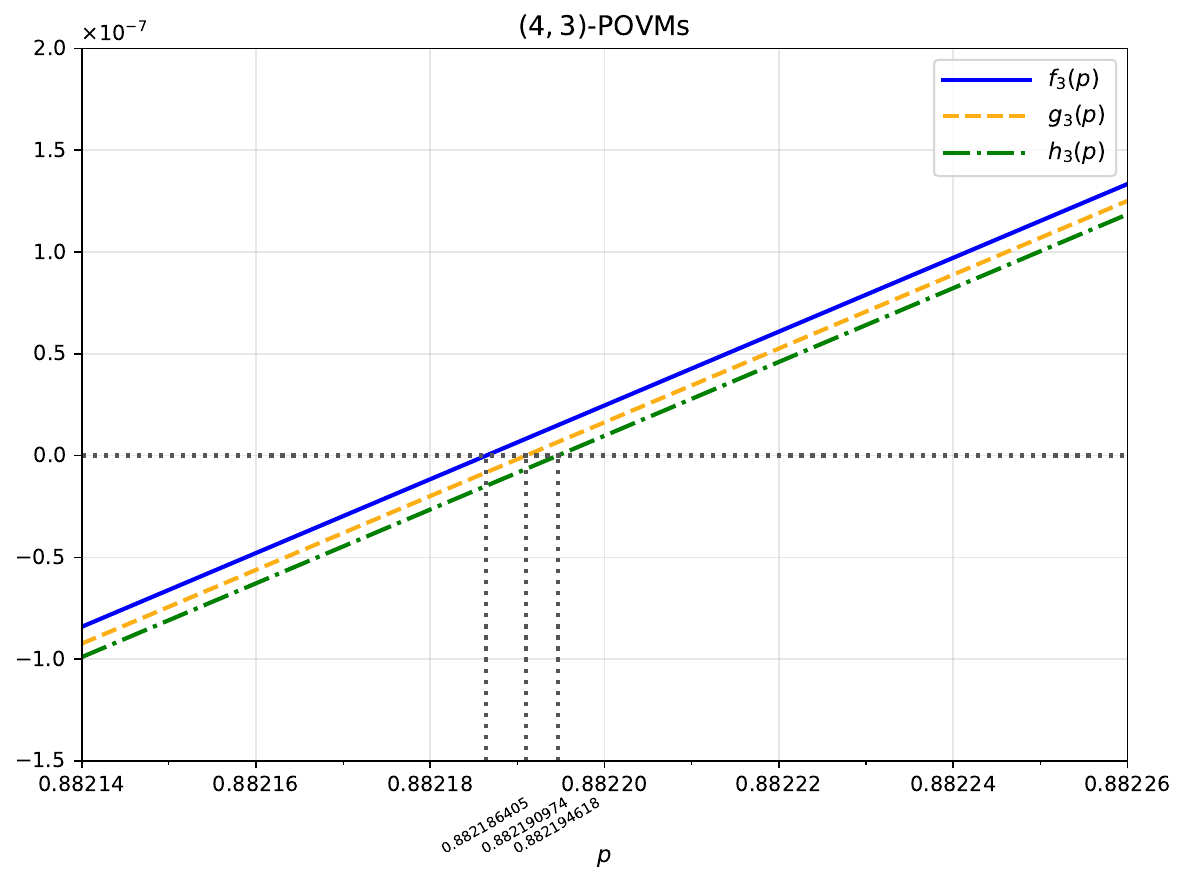}
	\caption{$f_3(p)$ derived from Corollary \ref{cor2} (solid blue line). $g_3(p)$ from Corollary 2 in Ref. \cite{LuAQTe70344} (dashed orange line). $h_3(p)$ presented in Corollary 2 in Ref. \cite{LuPRA062424} (dot-dashed green line).}
	\label{example1fig3}
\end{figure}
\end{example}

\section{Lower bounds of concurrence}\label{sec:concurrence}

In this section, we show that the general matrix $\mathcal{Q}_{P,Q,\mu,\nu}(\rho_{AB})$ also gives rise to an improved lower bound of the concurrence. Recall that the concurrence of a pure state $|\psi\rangle\in\mathcal{H}_A\otimes\mathcal{H}_B$ is defined by
\begin{equation}\label{PureConcurrence}
C(|\psi\rangle)=\sqrt{2\Big[1-\mathrm{Tr}\Big(\rho_A^{2}\Big)\Big]},
\end{equation}
where $\rho_A=\mathrm{Tr}_B(|\psi\rangle\langle\psi|)$ is the reduced state over subsystem $A$. For a mixed state $\rho$, its concurrence is defined by the convex roof extension,
\begin{equation}\label{MixedConcurrence}
C(\rho)=\min_{\{p_k,|\psi_k\rangle\}}\sum_k p_k C(|\psi_k\rangle),
\end{equation}
where the minimum is taken over all pure-state decompositions $\rho=\sum\limits_k p_k |\psi_k\rangle\langle\psi_k|$ with $p_k\geq0$ and $\sum\limits_k p_k = 1$.

\begin{theorem}\label{thmconcurrence}
For any bipartite state $\rho$  on $\mathcal{H}_A\otimes\mathcal{H}_B$, the concurrence $C(\rho)$ satisfies
\begin{equation}\label{ConcurrenceLowerBound}
C(\rho)\geq K\sqrt{\frac{2}{d(d-1)}\,}\Big(\left\|\mathcal{Q}_{P,Q,\mu,\nu}(\rho)\right\|_{\mathrm{tr}}-\mathfrak{M}(P,Q,\mu,\nu,M_A,M_B,d_A,d_B,x_A,x_B)\Big),
\end{equation}
where
$K=\sqrt{\dfrac{M_A(M_A-1)M_B(M_B-1)}{\left(M_A^{2}x_A-d_A\right)\left(M_B^{2}x_B-d_B\right)}\,}$
and $d=\min\{d_A,d_B\}$.
\end{theorem}

\begin{proof}
Let $\{p_k,|\psi_k\rangle\}$ be the optimal pure-state decomposition of $\rho$ such that $C(\rho)=\sum_k p_k C(|\psi_k\rangle)$. Since the map $\rho\mapsto\mathcal{Q}_{P,Q,\mu,\nu}(\rho)$ is linear and the trace norm is convex,
\begin{equation*}
\sum_k p_k \left\|\mathcal{Q}_{P,Q,\mu,\nu}(|\psi_k\rangle\langle\psi_k|)\right\|_{\mathrm{tr}}
\geq\left\|\sum_k p_k \mathcal{Q}_{P,Q,\mu,\nu}(|\psi_k\rangle\langle\psi_k|)\right\|_{\mathrm{tr}}
=\left\|\mathcal{Q}_{P,Q,\mu,\nu}(\rho)\right\|_{\mathrm{tr}}.
\end{equation*}
It suffices to prove the theorem for pure states \(\ket{\psi}\in\mathcal{H}_A\otimes\mathcal{H}_B\). That is,
\begin{equation}\label{PureStateSituation}
C(|\psi\rangle\langle\psi|)\geq K\sqrt{\frac{2}{d(d-1)}}\left(\left|\mathcal{Q}{P,Q,\mu,\nu}(|\psi\rangle\langle\psi|)\right|{\mathrm{tr}}-\mathfrak{M}(P,Q,\mu,\nu,M_A,M_B,d_A,d_B,x_A,x_B)\right).
\end{equation}

By the Schmidt decomposition, there exist orthonormal sets \(\{\ket{e_k}\}_{k=1}^d\) of \(\mathcal{H}_A\) and \(\{\ket{f_k}\}_{k=1}^d\) of \(\mathcal{H}_B\), respectively, such that
\(\ket{\psi}=\sum_{k=1}^{d}\lambda_k|e_k f_k\rangle,\)
where \(\lambda_k> 0\) and \(\sum_{k=1}^{d}\lambda_k^2=1\). Denote the separable state
$\xi=\sum_{k=1}^{d}\lambda_k^2|e_kf_k\rangle\langle e_kf_k|$. Obviously, $\mathrm{Tr}_B(\xi)=\mathrm{Tr}_B(|\psi\rangle\langle\psi|)$ and $\mathrm{Tr}_A(\xi)=\mathrm{Tr}_A(|\psi\rangle\langle\psi|)$. Hence, the probability vectors $\tau$ and $\sigma$ defined in Eq.~(\ref{SubsystemsMeanmentResults}) coincide for $\xi$ and $|\psi\rangle\langle\psi|$. Consequently,
\begin{equation*}
\mathcal{Q}_{P,Q,\mu,\nu}(|\psi\rangle\langle\psi|)=\mathcal{Q}_{P,Q,\mu,\nu}(\xi)
+\begin{pmatrix}
0 & 0 \\[2pt]
0 & \mathcal{P}(|\psi\rangle\langle\psi|)-\mathcal{P}(\xi)
\end{pmatrix}.
\end{equation*}
Thus, we have
\begin{equation}\label{SplitBound}
\left\|\mathcal{Q}_{P,Q,\mu,\nu}(|\psi\rangle\langle\psi|)\right\|_{\mathrm{tr}}
\leq\left\|\mathcal{Q}_{P,Q,\mu,\nu}(\xi)\right\|_{\mathrm{tr}}
+\left\|\mathcal{P}(|\psi\rangle\langle\psi|)-\mathcal{P}(\xi)\right\|_{\mathrm{tr}}.
\end{equation}

Since  $\xi$ is separable, Theorem \ref{thm1} yields
\begin{equation*}
\left\|\mathcal{Q}_{P,Q,\mu,\nu}(\xi)\right\|_{\mathrm{tr}}\leq\mathfrak{M}(P,Q,\mu,\nu,d_A,d_B,M_A,M_B,x_A,x_B).
\end{equation*}

It remains to evaluate the second term of Eq.~(\ref{SplitBound}). Let $\{|g_{\alpha,k}\rangle\mid\alpha=1,\cdots,N_A;\,k=1,\cdots,M_A\}$ and $\{|h_{\beta,\ell}\rangle\mid\beta=1,\cdots,N_B;\,\ell=1,\cdots,M_B\}$ be orthonormal bases of $\mathbb{C}^{N_AM_A}$ and $\mathbb{C}^{N_BM_B}$, respectively. Define $\mathcal{P}'(\rho)=\sum_{\alpha=1}^{N_A}\sum_{k=1}^{M_A}\sum_{\beta=1}^{N_B}\sum_{\ell=1}^{M_B}\mathrm{Tr}\Big[\left(E^{A}_{\alpha,k}\otimes E^{B}_{\beta,\ell}\right)\rho\Big]|g_{\alpha,k}\rangle\langle h_{\beta,\ell}|$,
$|u_{ij}\rangle=\sum_{\alpha=1}^{N_A}\sum_{k=1}^{M_A}\langle e_j|E^{A}_{\alpha,k}|e_i\rangle|g_{\alpha,k}\rangle$,
$|v_{ij}\rangle=\sum_{\beta=1}^{N_B}\sum_{\ell=1}^{M_B}\langle f_j|E^{B}_{\beta,\ell}|f_i\rangle|h_{\beta,\ell}\rangle$.
Then
\begin{equation*}
\begin{aligned}
\mathcal{P}'(|\psi\rangle\langle\psi|)&=\sum_{\alpha=1}^{N_A}\sum_{k=1}^{M_A}\sum_{\beta=1}^{N_B}\sum_{\ell=1}^{M_B}\sum_{i=1}^{d}\sum_{j=1}^{d}\lambda_i\lambda_j\langle e_j|E^A_{\alpha,k}|e_i\rangle\langle f_j|E^B_{\beta,\ell}|f_i\rangle\,|g_{\alpha,k}\rangle\langle h_{\beta,\ell}| \\
&=\sum_{i=1}^{d}\sum_{j=1}^{d}\lambda_i\lambda_j|u_{ij}\rangle\langle\overline{v}_{ij}|,
\end{aligned}
\end{equation*}
where $|\overline{v}_{ij}\rangle=\sum_{\beta=1}^{N_B}\sum_{\ell=1}^{M_B}\overline{\langle f_j|E^B_{\beta,\ell}|f_i\rangle}|h_{\beta,\ell}\rangle$ with $\overline{\langle f_j|E^B_{\beta,\ell}|f_i\rangle}$ standing for the complex conjugation of $\langle f_j|E^B_{\beta,\ell}|f_i\rangle$. And $\mathcal{P}'(\xi)=\sum_{k=1}^{d}\lambda_k^2|u_{kk}\rangle\langle\overline{v}_{kk}|$. Therefore, we have
\begin{equation*}
\|\mathcal{P}(|\psi\rangle\langle\psi|)-\mathcal{P}(\xi)\|_{\mathrm{tr}}=
\|\mathcal{P}^{\prime}(|\psi\rangle\langle\psi|)-\mathcal{P}^{\prime}
(\xi)\|_{\mathrm{tr}}=\left\|\sum\limits_{i\neq j}\lambda_{i}\lambda_{j}\ket{u_{ij}}\bra{\bar{v}_{ij}}\right\|_{\mathrm{tr}},
\end{equation*}
where the first equality follows from $\mathcal{P}^{\prime}
=U\left(\sum_{\alpha=1}^{N_A}\sum_{k=1}^{M_A}\sum_{\beta=1}^{N_B}\sum_{\ell=1}^{M_B}\mathrm{Tr}\Big[\left(E^{A}_{\alpha,k}\otimes E^{B}_{\beta,\ell}\right)\rho\Big]|\alpha,k\rangle\langle \beta,\ell|\right)V^{\dagger}
=U\mathcal{P}V^{\dagger}$ and the fact that the trace norm is invariant under isometric transformations, with $U$ and $V$ satisfying $|g_{\alpha,k}\rangle=U|\alpha,k\rangle$ and $|h_{\beta,\ell}\rangle=V|\beta,\ell\rangle$, respectively.

We denote $F=\sum\limits_{i=1}^{d}\sum\limits_{j=1}^{d}\ket{e_{i}}\bra{e_{j}}\otimes\ket{e_{j}}\bra{e_{i}}$. Since $\sum\limits_{\alpha=1}^{N_{A}}\sum\limits_{k=1}^{M_{A}}E_{\alpha,k}^{A}\otimes E_{\alpha,k}^{A}=\dfrac{M_{A}^{2}x_{A}-d_{A}}{M_{A}(M_{A}-1)}F+\dfrac{d_{A}^{3}-M_{A}^{2}x_{A}}{d_{A}M_{A}(M_{A}-1)}\mathbb{I}_A\otimes\mathbb{I}_A$ \cite{SMdesign1,SMdesign2}, we have
\begin{align}
\langle u_{i^{\prime}j^{\prime}}|u_{ij}\rangle&=\sum\limits_{\alpha,\beta=1}^{N_A}\sum\limits_{k,\ell=1}^{M_A}\overline{\bra{e_{j^{\prime}}}E^A_{\alpha,k}|e_{i^{\prime}}\rangle}\bra{e_{j}}E^A_{\beta,\ell}|e_{i}\rangle\langle g_{\alpha,k}|g_{\beta,\ell}\rangle \nonumber\\
&=\sum\limits_{\alpha=1}^{N_A}\sum\limits_{k=1}^{M_A}\bra{e_{i^{\prime}}}E^A_{\alpha,k}|e_{j^{\prime}}\rangle\bra{e_{j}}E^A_{\alpha,k}|e_{i}\rangle \nonumber\\
&=\sum\limits_{\alpha=1}^{N_A}\sum\limits_{k=1}^{M_A}\bra{e_{i^{\prime}}e_{j}}E^A_{\alpha,k}\otimes E^A_{\alpha,k}\ket{e_{j^{\prime}}e_{i}} \nonumber\\
&=\bra{e_{i^{\prime}}e_{j}}\left(\sum\limits_{\alpha=1}^{N_A}\sum\limits_{k=1}^{M_A}E_{\alpha,k}\otimes E_{\alpha,k}\right)|e_{j^{\prime}}e_{i}\rangle \nonumber\\
&=\dfrac{M_A^{2}x_A-d_A}{M_A(M_A-1)}\bra{e_{i^{\prime}}e_{j}}F|e_{j^{\prime}}e_{i}\rangle+\dfrac{d_A^{3}-M_A^{2}x_A}{d_AM_A(M_A-1)}\bra{e_{i^{\prime}}e_{j}}e_{j^{\prime}}e_{i}\rangle \nonumber\\
&=\dfrac{M_A^{2}x_A-d_A}{M_A(M_A-1)}\delta_{i^{\prime}i}\delta_{j^{\prime}j}+\dfrac{d_A^{3}-M_A^{2}x_A}{d_AM_A(M_A-1)}\delta_{ij}\delta_{i^{\prime}j^{\prime}} \label{braket-u}.
\end{align}
Similarly, we have $\bra{v_{i^{\prime}j^{\prime}}}v_{ij}\rangle=\dfrac{M_{B}^{2}x_{B}-d_{B}}
{M_{B}(M_{B}-1)}\delta_{i^{\prime}i}\delta_{j^{\prime}j}+\dfrac{d_{B}^{3}-M_{B}^{2}x_{B}}
{d_{B}M_{B}(M_{B}-1)}\delta_{ij}\delta_{i^{\prime}j^{\prime}}$. Noting that $\bra{\overline{v}_{i^{\prime}j^{\prime}}}\overline{v}_{ij}\rangle
=\overline{\langle v_{i^{\prime}j^{\prime}}|v_{ij}\rangle}$, we obtain
\begin{equation*}
\bra{\overline{v}_{i^{\prime}j^{\prime}}}\overline{v}_{ij}\rangle
=\dfrac{M_{B}^{2}x_{B}-d_{B}}{M_{B}(M_{B}-1)}\delta_{i^{\prime}i}
\delta_{j^{\prime}j}+\dfrac{d_{B}^{3}-M_{B}^{2}x_{B}}{d_{B}M_{B}(M_{B}-1)}
\delta_{ij}\delta_{i^{\prime}j^{\prime}}.
\end{equation*}
Therefore, both $\{\ket{u_{ij}}|i\neq j\}$ and $\{\ket{\overline{v}_{ij}}|i\neq j\}$ are orthogonal sets. Write $\mathcal{Q}=\sum_{i\neq j}\lambda_{i}\lambda_{j}\ket{u_{ij}}\bra{\bar{v}_{ij}}$, then $\mathcal{Q}^{\dagger}\mathcal{Q}=\sum_{i\neq j}(\lambda_{i}\lambda_{j})^{2}\bra{u_{ij}}u_{ij}\rangle
\ket{\bar{v}_{ij}}\bra{\bar{v}_{ij}}$. Hence, we have that
\begin{equation*}
\|\mathcal{P}(|\psi\rangle\langle\psi|)-\mathcal{P}(\xi)\|_{\mathrm{tr}}=\|\mathcal{Q}\|_{\mathrm{tr}}=\sum_{i\neq j}\lambda_{i}\lambda_{j}\sqrt{\bra{u_{ij}}u_{ij}\rangle}\sqrt{\bra{\bar{v}_{ij}}\bar{v}_{ij}\rangle}=\frac{2}{K}\sum_{i<j}\lambda_{i}\lambda_{j}, 
\end{equation*}
where the last equality is from Eqs.~(\ref{braket-u}).

Consequently, we have
\begin{equation}\label{PureStatesBound}
\left\|\mathcal{Q}_{P,Q,\mu,\nu}(|\psi\rangle\langle\psi|)\right\|_{\mathrm{tr}}
\leq\mathfrak{M}(P,Q,\mu,\nu,d_A,d_B,M_A,M_B,x_A,x_B)+\frac{2}{K}\sum_{i < j}\lambda_i\lambda_j.
\end{equation}
On the other hand, it is known that \cite{ChenAlbeverioFei}
\begin{equation*}
C(|\psi\rangle)\geq2\sqrt{\frac{2}{d(d-1)}}\sum_{i<j}\lambda_i\lambda_j,
\end{equation*}
which gives rise to Eq.~(\ref{PureStateSituation}) by using Eq.~(\ref{PureStatesBound}).
\end{proof}

\textit{Remark 1.} If $\rho$ is a separable state, i.e. $C(\rho)=0$, then (\ref{ConcurrenceLowerBound}) gives rise to that $$\|\mathcal{Q}_{P,Q,\mu,\nu}(\rho_{AB})\|_{\mathrm{tr}}\leq\mathfrak{M}(P,Q,\mu,\nu,M_A,M_B,d_A,d_B,x_A,x_B),$$ which is consistent with the Theorem \ref{thm1}.

\textit{Remark 2.} Taking $P=Q=0$ and $\mu=\nu=1$ in Theorem \ref{thmconcurrence}, we have
\begin{equation*}
C(\rho) \geq K\sqrt{\dfrac{2}{d(d-1)}}\left(\|\mathcal{P}(\rho)\|_{\rm tr}-\sqrt{\dfrac{(d_{A}-1)(M_{A}^{2}x_{A}+d_{A}^{2})}{d_{A}M_{A}(M_{A}-1)}}\sqrt{\dfrac{(d_{B}-1)(M_{B}^{2}x_{B}+d_{B}^{2})}{d_{B}M_{B}(M_{B}-1)}}\,\right).
\end{equation*}
Hence, Theorem \ref{thmconcurrence} generalizes the lower bounds of concurrence based on symmetric measurements proposed in Ref. \cite{WangFeiConcurrence}. If we take $P=\boldsymbol{a},Q=\boldsymbol{b}$ and $\mu=\nu=1$ in Theorem \ref{thmconcurrence}, then 
\begin{equation*}
C(\rho) \geq K\sqrt{\dfrac{2}{d(d-1)}}\left(\|\mathcal{Q}_{\bm{a},\bm{b}}(\rho)\|_{\mathrm{tr}}-\sqrt{|\bm{a}|^{2}+\dfrac{(d_{A}-1)(M_{A}^{2}x_{A}+d_{A}^{2})}{d_{A}M_{A}(M_{A}-1)}}\sqrt{|\bm{b}|^{2}+\dfrac{(d_{B}-1)(M_{B}^{2}x_{B}+d_{B}^{2})}{d_{B}M_{B}(M_{B}-1)}}\,\right).
\end{equation*}
Therefore, Theorem \ref{thmconcurrence} generalizes the lower bounds of concurrence based on symmetric measurements proposed in Ref. \cite{WangEtAlEstimatingConcurrence}. 

To show the advantages of Theorem \ref{thmconcurrence}, we give the following example.

\begin{figure}[htbp]
	\includegraphics[width=0.8\textwidth]{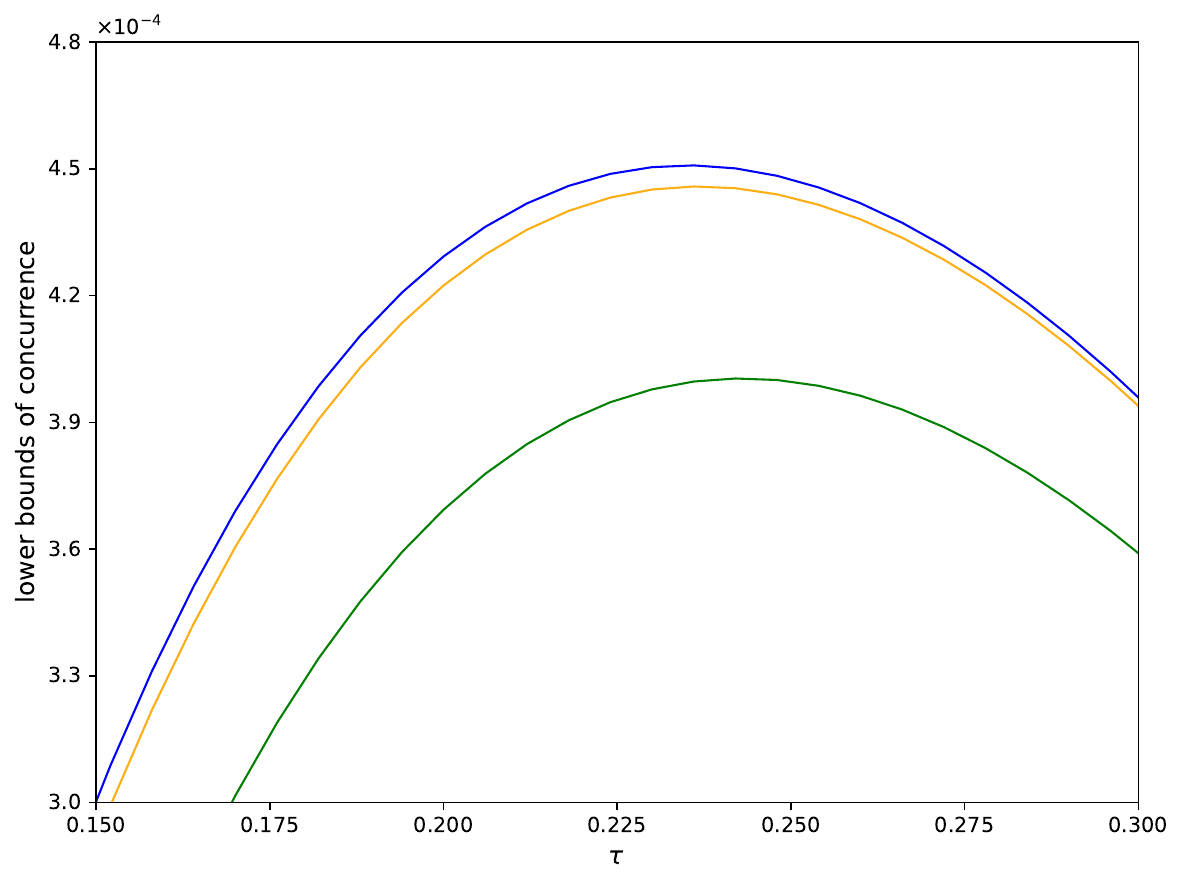}
	\caption{Lower bounds of $C(\rho(\tau,q))$. The blue line for the bound from our Theorem \ref{thmconcurrence}, the orange line for the bound from Ref.~\cite{WangEtAlEstimatingConcurrence}, and the green line for the bound from Ref.~\cite{WangFeiConcurrence}.}
	\label{concurrence}
\end{figure}

\begin{example}
Consider the following mixture of the bound entangled state proposed by Horodecki \cite{HorodeckiPLA333},
	\begin{equation*}
		\rho_{\tau}=\dfrac{1}{1+8\tau}
		\begin{pmatrix}
			\tau & 0 & 0 & 0 & \tau & 0 & 0 & 0 & \tau \\
			0 & \tau & 0 & 0 & 0 & 0 & 0 & 0 & 0 \\
			0 & 0 & \tau & 0 & 0 & 0 & 0 & 0 & 0 \\
			0 & 0 & 0 & \tau & 0 & 0 & 0 & 0 & 0 \\
			\tau & 0 & 0 & 0 & \tau & 0 & 0 & 0 & \tau \\
			0 & 0 & 0 & 0 & 0 & \tau & 0 & 0 & 0 \\
			0 & 0 & 0 & 0 & 0 & 0 & \frac{1+\tau}{2} & 0 & \frac{\sqrt{1-\tau^{2}}}{2}\\
			0 & 0 & 0 & 0 & 0 & 0 & 0 & \tau & 0 \\
			\tau & 0 & 0 & 0 & \tau & 0 & \frac{\sqrt{1-\tau^{2}}}{2} & 0 & \frac{1+\tau}{2}
		\end{pmatrix}
	\end{equation*}
and the $9 \times 9$ identity matrix $I_{9}$,
\begin{equation*}
	\rho(\tau,q)=\frac{1-q}{9}I_{9}+q\rho_{\tau}.
\end{equation*}

FIG. \ref{concurrence} illustrates the lower bounds of $C(\rho(\tau,q))$ for $q=0.995$. The blue curve is the bound from Theorem 1 in Ref.~\cite{WangFeiConcurrence} based on the $(8,2)$-POVMs in Example \ref{eg1} with $t=0.01$. The orange curve is the bound from Theorem 1 in Ref.~\cite{WangEtAlEstimatingConcurrence} for $\boldsymbol{a}=(0.47,-0.91)^T$, $\boldsymbol{b}=(-0.97,-0.02)^T$ and the $(8,2)$-POVMs in Example \ref{eg1} with $t=0.01$. The green curve is the lower bound from our Theorem \ref{thmconcurrence} with the $(8,2)$-POVMs in Example \ref{eg1} and
\begin{equation*}
P=
\begin{pmatrix}
-0.82 & 0.53 \\
-2.59 & 0.38 \\
-1.45 & 1.5 \\
-0.98 & -0.46 \\
0.31 & 1.86
\end{pmatrix},\quad
Q=
\begin{pmatrix}
0.2 & -1.11 \\
0.16 & 0.93 \\
0.05 & -0.55 \\
2.19 & 0.78 \\
0.71 & -0.36 \\
-0.7 & 0.58
\end{pmatrix},\quad
\mu=
	\begin{pmatrix}
		\frac{1}{\sqrt{2}} \\[3pt]
		\frac{1}{\sqrt{2}}
	\end{pmatrix},\quad
	\nu=
	\begin{pmatrix}
		\frac{1}{\sqrt{2}} \\[3pt]
		\frac{1}{\sqrt{2}}
	\end{pmatrix}.
\end{equation*}

\end{example}

\section{Conclusions}\label{sec5}

We have presented a new separability criteria by constructing matrices via realignment of measurement probabilities arising from symmetric measurements. The key novelty lies in the fact that the auxiliary block in the upper-left corner of the constructed matrix \(\mathcal{Q}_{P,Q,\mu,\nu}(\rho)\) takes the form of an arbitrary-ranked matrix \(PQ^T\), in contrast to the scalars or rank-one matrices adopted in prior works. Within this framework, our matrix recovers the constructions put forward by Lu \textit{et al.} as special cases. The associated separability criterion offers substantially greater flexibility, with all the related earlier criteria being recovered by setting \(r=1\). Furthermore, we have derived lower bounds on the concurrence of bipartite states by using this matrix construction. Detailed examples show that our criteria detect more entangled states and provide tighter lower bounds of concurrence than the existing benchmarks.

Recently, K. Siudzi\'nska introduced generalized equiangular weighted frames (GEWFs) \cite{Siudzinska2026JPA445302}, which possess crucial symmetry properties and for which the index of coincidence admits an analytical upper bound. Therefore, using GEWFs to construct an association matrix about the measurement results to obtain a more powerful entanglement criterion and the lower bound of the Schmidt number is a potential research direction.

\section*{Acknowledgments:}
S.-M. Fei acknowledges financial support from the Specific Research Fund of the Innovation Platform for Academicians of Hainan Province.

\section*{Data availability}
No data were created or analyzed in this study.

\end{document}